\documentclass[aps,prl,twocolumn,showpacs,groupaddress]{revtex4}
\usepackage{epsfig}
\usepackage{bm}
\begin{document}
\title{
Imaging quasi-particle wavefunctions in quantum dots 
via tunneling spectroscopy 
}

\author{Massimo Rontani}

\author{Elisa Molinari}

\affiliation{INFM National Research Center on nanoStructures
and bioSystems at Surfaces (S3), and  Dipartimento di Fisica,\\
Universit\`a degli Studi di Modena e Reggio Emilia,
Via Campi 213/A, 41100 Modena, Italy}

\date{\today}
\begin{abstract}
We show that in quantum dots the physical quantities probed by local 
tunneling spectroscopies, namely the quasi-particle wavefunctions of 
interacting electrons, can considerably deviate from their single-particle 
counterparts as an effect of Coulomb correlation. From the exact solution 
of the few-particle Hamiltonian for prototype dots,
we find that such deviations are crucial to predict wavefunction images 
at low electron densities or high magnetic fields.
\end{abstract}
\pacs{73.21.La, 73.23.Hk, 73.20.Qt, 73.63.Kv}

\maketitle

Current single-electron tunneling spectroscopies in semiconductor
quantum dots (QDs) \cite{Jacak,Bimberg,Reimann}
may provide spectacular images of the QD wavefunctions,
both in real \cite{Grandidier,Millo,Maltezopoulos}
and reciprocal space \cite{Vdovin,Patane,Wibbelhoff}.
The measured intensities have been generally attributed 
to the probability densities of
ground or excited single electron states occupying the dot.
As pointed out by Wibbelhoff {\em et al.} \cite{Wibbelhoff}, however,
the role of other electrons filling the dot may actually be relevant.
Indeed, QDs can be strongly interacting objects with a
completely discrete energy spectrum, which in turn depends on the
number of electrons, $N$ \cite{Jacak,Reimann}. 
Therefore, orbitals can be ill-defined,
losing their meaning due to interaction. Also, it is
unclear how many electrons one should take into account
to calculate the total density of states, 
as a particle tunnels into a QD filled with $N$ electrons.
In this Letter we thus address the following basic questions: What are
the physical quantities that are actually probed by scanning tunneling
microscopies (STM) \cite{Grandidier,Millo,Maltezopoulos}
or magneto-tunneling spectroscopies \cite{Vdovin,Patane,Wibbelhoff} of QDs?
How do they depend on interactions? Can they deviate from the common
single-particle picture in physically relevant regimes?
If only one many-body state is probed at time,
then the signal is proportional to the probability density
of the {\em quasi-particle} (QP) being injected into the interacting QD.
We demonstrate that the QP density dramatically
depends on the strength of correlation inside the dot, 
and predict the wavefunction mapping to be a useful experimental tool
to image QPs, both in direct and reciprocal space. 

The imaging experiments, in their essence, measure
quantities directly proportional to the 
probability for transfer of an electron through a barrier, from an  
emitter, where electrons fill in a Fermi sea, to a dot, with
completely discrete energy spectrum. 
In multi-terminal setups one can neglect the role of electrodes 
other than the emitter, to a first approximation.
The measured quantity can be
the current \cite{Grandidier,Vdovin}, the differential conductance 
\cite{Millo,Patane,Maltezopoulos,Seigo}, or the QD capacitance 
\cite{Wibbelhoff,Ashoori}, while the emitter can be the STM tip 
\cite{Grandidier,Millo,Maltezopoulos}, or a $n$-doped GaAs contact
\cite{Vdovin,Patane,Wibbelhoff,Seigo,Ashoori}, 
and the barrier can be the vacuum
\cite{Grandidier,Millo,Maltezopoulos} as well as a AlGaAs spacer
\cite{Vdovin,Patane,Wibbelhoff,Seigo,Ashoori}. 

According to the seminal paper by Bardeen \cite{Bardeen},
the transition probability (at zero temperature) is given by the expression
$(2\pi/\hbar)\left|{\cal{M}}\right|^2n(\epsilon_f)$, where
$\cal{M}$ is the matrix element and $n(\epsilon_f)$ is the energy
density of the final QD states. The common wisdom would predict the
probability to be proportional to the total density of QD states
at the resonant tunneling energy, $\epsilon_f$, possibly space-resolved 
since $\cal{M}$ would depend on the resonant QD orbital \cite{Tersoff}. 
To proceed, let us assume that electrons from
the emitter access through the barrier a single QD at a sharp resonant energy, 
corresponding to a unique, well defined many-body QD state,
and reconsider the transition
matrix element \protect{${\cal{M}}_{\bm{k},N}$} for transfer of an electron 
from emitter to QD. 
\protect{${\cal{M}}_{\bm{k},N}$} is given by [recasting Eqs.~(6-7) 
of Ref.~\onlinecite{Bardeen}
in second quantized form]:
\begin{eqnarray}
&& {\cal{M}}_{\bm{k},N}  \propto  \langle \{k\},N-1 |
\,\hat{\cal{M}}\,
|\{k^*\},N \rangle , \nonumber\\
&& \hat{\cal{M}}  =  \frac{\hbar^2}{2m^*}\int\left[{\hat{\Psi}}^{\dagger}
\frac{\partial \hat{\Psi} }{\partial z} - \frac{\partial 
{\hat{\Psi}}^{\dagger} }{\partial z} \hat{\Psi}\right]
\delta(z_{\text{bar}}-z)\,\text{d}\,\tau.
\label{eq:bardeen}
\end{eqnarray}
Here $| \{k\},N-1 \rangle$ and $|\{k^*\},N \rangle$ 
are two many-particle states
of the entire system of similar energies, with $N-1$ and
$N$ interacting electrons in the QD, respectively, and the remaining
$N_{\text{tot}} - N + 1$ and $N_{\text{tot}} - N$ electrons, 
respectively, in the emitter.
The fixed coordinate along the tunneling direction $z$ appearing in
(\ref{eq:bardeen}), $z_{\text{bar}}$, can be
everywhere in the barrier, and $\text{d}\,\tau$ is the infinitesimal
volume element. 
The fermionic field operator $\hat{\Psi}(\bm{r})$, destroying an electron
at position $\bm{r}\equiv (\bm{\varrho},z)$, can be expanded over the
basis of emitter and QD single-particle states, $\phi_{\bm{k}}$
and $\phi_{\alpha}$, respectively \cite{orthogonality}:
\protect{$\hat{\Psi}(\bm{r})=\sum_{i}\phi_i(\bm{r}){\hat{c}}_i$},
where \protect{$i=\bm{k},\alpha$} and we 
take unitary volume normalization. We omit spin indexes and
summations for the sake of simplicity.
We assume that electrons in the emitter do not interact and
are associated to the two sets of quantum numbers $\{k\}$ and $\{k^*\}$,
respectively, which differ in the occurrence of
the index $\bm{k}$ labeling the electron which leaves the emitter and
tunnels to the QD as
$|\{k\},N-1\rangle$ evolves to $|\{k^*\},N\rangle$.
Moreover, we assume for convenience that the $xy$ and $z$ motions of electrons
are separable, and that electrons
in the QD all occupy the same confined single-particle state along $z$,
$\chi_{\text{QD}}(z)$, namely $\phi_{\alpha}(\bm{r}) = 
\phi_{\alpha}(\bm{\varrho})\chi_{\text{QD}}(z)$.
Under these conditions we may factorize
the matrix element as 
\protect{${\cal{M}}_{\bm{k},N}\propto T_{\bm{k}}\,M_{\bm{k},N}$}, with
\begin{equation}
T_{\bm{k}} = \frac{\hbar^2}{2m^*}\left[\chi^*_{\bm{k}}(z)
\frac{ \partial \chi_{\text{QD}}(z) }{\partial z}
-\chi_{\text{QD}}(z)\frac{\partial \chi_{\bm{k}}^*(z) }{\partial z}
\right]_{ z=z_{\text{bar}} },
\label{eq:T}
\end{equation}
where $\chi_{\bm{k}}(z)$ is the emitter state along $z$ evanescent 
in the barrier,
\protect{$\phi_{\bm{k}}(\bm{r})=\phi_{\bm{k}}(\bm{\varrho})\,
\chi_{\bm{k}}(z)$},
%\protect{$\phi_{\bm{k}}(\bm{r})=\text{e}^{\text{i}\bm{k\cdot\varrho}}
%\chi_{\bm{k}}(z)$}, 
and
\begin{displaymath}
M_{\bm{k},N} = \sum_{\alpha} \!\int\!
\phi^*_{\bm{k}}(\bm{\varrho})\, \phi_{\alpha}\!(\bm{\varrho})\,
\text{d}\,\bm{\varrho} \,\langle \{k\},N-1|\,{\hat{c}}^{\dagger}_{\bm{k}}
{\hat{c}}_{\alpha} |\{k^*\},N \rangle.
\end{displaymath}
Eventually assuming that the many-body states can be factorized
into an emitter and a QD part, we obtain
\begin{equation}
M_{\bm{k},N} = \int \phi^*_{\bm{k}}(\bm{\varrho})\,
\varphi_{\text{QD}}(\bm{\varrho})\,\text{d}\,\bm{\varrho},
\label{eq:M}
\end{equation}
where $\varphi_{\text{QD}}(\bm{\varrho})$ is the quasi-particle (QP)
wavefunction of the interacting QD system \cite{nota}:
\begin{equation}
\varphi_{\text{QD}}(\bm{\varrho}) = \langle N - 1 | \hat{\Psi}(\bm{\varrho})|
N \rangle .
\label{eq:def}
\end{equation}

Results (\ref{eq:M}-\ref{eq:def}) are the key
for predicting  wavefunction images both in real
and reciprocal space. 
In STM, $\phi_{\bm{k}}(\bm{\varrho})$ is the localized tip
wavefunction; if we ideally assume it point-like and
located at $\bm{\varrho}_0$ \cite{Tersoff},
i.e.~\protect{$\phi_{\bm{k}}(\bm{\varrho}) \approx \delta(\bm{\varrho}
-\bm{\varrho}_0)$}, then the signal intensity is proportional
to $\left|\varphi_{\text{QD}}(\bm{\varrho}_0)\right|^2$,
which is the usual result of the one-electron
theory \cite{Maltezopoulos,Tersoff}, provided the ill-defined QD orbital
is replaced by the QP wavefunction unambiguously defined by Eq.~(\ref{eq:def}).
In magneto-tunneling spectroscopy, the emitter in-plane wavefunction
is a plane wave, \protect{$\phi_{\bm{k}}(\bm{\varrho})=
\text{e}^{\text{i}\bm{k\cdot\varrho}}$}, and the matrix element
(\ref{eq:M}) is the Fourier transform of $\varphi_{\text{QD}}$,
\protect{$M_{\bm{k},N} = \varphi_{\text{QD}}(\bm{k})$}.
Again, we generalize the standard one-electron result by
substituting $\varphi_{\text{QD}}(\bm{k})$ for
the QD orbital [then Eqs.~(\ref{eq:M}-\ref{eq:T}) coincide with (A1-A2) of
Ref.~\onlinecite{Patane}].
Note that $M_{\bm{k},N}$ is the relevant quantity 
also for intensities in space-integrated spectroscopies probing the QD
addition energy spectrum \cite{Seigo,Ashoori}. 
Consistently, in the non-interacting case,
$\varphi_{\text{QD}}(\bm{\varrho})$ reduces 
to the highest occupied one-electron orbital 
$\phi_{\alpha}(\bm{\varrho})$ \cite{Tersoff}:
in this limit an electron tunnels from the emitter to the orbital
$\phi_{\alpha}(\bm{\varrho})$ which resonates at the Fermi energy, with
$|N\rangle = {\hat{c}}^{\dagger}_{\alpha}|N-1\rangle$.
The latter regime probably corresponds to most of the existing
experimental evidence 
\cite{Grandidier,Millo,Maltezopoulos,Vdovin,Patane,Wibbelhoff}.
However, it is interesting to analyze realistic scenarios that deviate
from the one-electron picture.

Therefore, we study $\varphi_{\text{QD}}(\bm{\varrho})$ in a paradigmatic 
interacting case, and consider a few electrons in a two-dimensional 
harmonic trap,
which was proven to be an excellent model for different experimental
setups \cite{Reimann}. The QD effective-mass Hamiltonian is
\begin{equation}
H = \sum_{i}^{N}H_{0}(i)+\frac{1}{2}\sum_{i\neq j}\frac{e^{2}}
{\kappa|\bm{\varrho}_{i}-\bm{\varrho}_{j}|},
\label{eq:HI}
\end{equation}
with
\begin{equation}
H_{0}(i)\,=\,\frac{1}{2m^{*}}\Big[\bm{p}-\frac{e}{c}
\bm{A}(\bm{\varrho})\Big]^{2}\,+\,m^*\omega_0^2\varrho^2/2.
\label{eq:HSP}
\end{equation}
Here $\kappa$ is the static relative dielectric constant of the host
semiconductor, and $\bm{A}(\bm{\varrho})$ is the vector
potential ($\bm{A}=\bm{B}\times\bm{\varrho}/2$)
associated with a static and uniform magnetic field $B$
along $z$, which reduces the cylindrical spatial symmetry group
of the system from $D_{\infty h}$, at $B=0$, to $C_{\infty h}$, when
$B\neq 0$, making it chiral. The QD wavefunction has an azimuthal
quantum number $m$, \protect{$\varphi_{\text{QD}}(\bm{\varrho})=
\varphi_{\text{QD}}(\varrho) \text{e}^{\text{i}m\varphi}$},
which is fixed by the total angular momenta $M$ of
$|N\rangle$ and $|N-1\rangle$, $m= M_N-M_{N-1}$, and can be expanded
over the basis of Fock-Darwin (FD) orbitals $\varphi_{nm}(\bm{\varrho})$
\cite{Jacak}, eigenstates of the single-particle Hamiltonian (\ref{eq:HSP}): 
\protect{$\varphi_{\text{QD}}(\bm{\varrho})=\sum_{n=0}^{\infty}
a_n\varphi_{nm}(\bm{\varrho})$}, where $n$'s are radial quantum numbers
and $a_n$ coefficients to be determined. We solve numerically the
few-body problem of Eq.~(\ref{eq:HI}), for the ground state at
different $N$'s, by means of
the configuration interaction (CI) method \cite{method}, where $|N\rangle$
is expanded in a series of Slater determinants built by filling in
the FD orbitals with $N$ electrons, and
consistently with symmetry constraints \cite{method}. 
Then, we evaluate the matrix
element (\ref{eq:def}), and find the values of $a_n$ for a truncated
FD basis set. 

There are two ways of artificially tuning the strength of Coulomb
correlation in QDs: one is to dilute electron density, and the other
is to turn on $B$. In both cases, at low enough densities or
strong enough fields, electrons pass from a ``liquid'' phase, 
where low-energy motion is equally controlled by kinetic and
Coulomb energy, to a ``crystallized'' phase, reminescent of the
Wigner crystal in the bulk, where electrons are localized in space
and arrange themselves in a geometrically ordered configuration
such that electrostatic repulsion is minimized \cite{Reimann,NATO}.

We first consider reducing the density at $B=0$. The typical 
QD lateral extension is given by the characteristic
dot radius $\ell_{\text{QD}} = (\hbar/m^*\omega_0)^{1/2}$, 
$\ell_{\text{QD}}$ being the
mean square root of $\varrho$ on the FD lowest-energy level $\varphi_{00}$.
As we keep $N$ fixed and increase $\ell_{\text{QD}}$, the Coulomb-to-kinetic
energy ratio $\lambda = \ell_{\text{QD}}/a^*_{\text{B}}$ [$a^*_{\text{B}}
=\hbar^2\kappa/(m^*e^2)$
is the effective Bohr radius of the dot] \cite{Egger} increases
as well, driving the system into the ``Wigner'' regime \cite{Brueckner}.
As a rough indication, consider that for $\lambda \approx 2$ or lower
the electronic ground state is liquid, while above $\lambda \approx 4$
electrons form a ``crystallized'' phase \cite{Egger}.
\begin{figure}
\setlength{\unitlength}{1 cm}
\begin{picture}(8.5,10)
\put(-0.3,-0.7){\epsfig{file=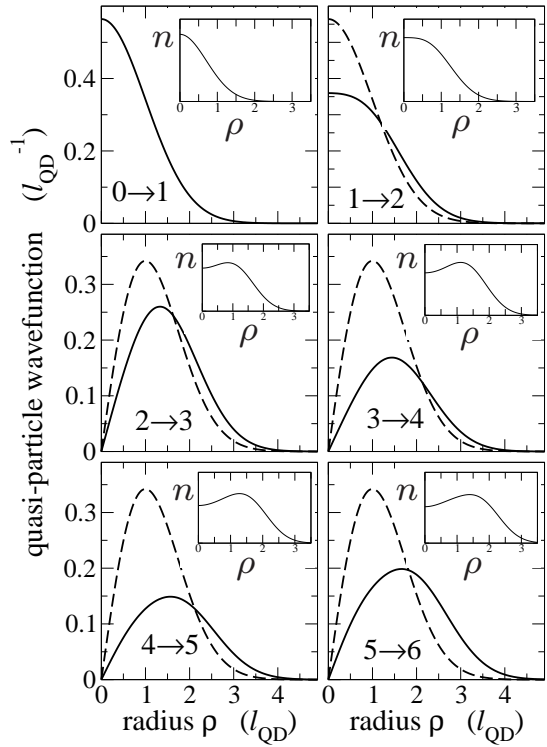,width=3.2in,angle=0}}
\end{picture}
\caption{Quasi-particle wavefunction (solid line) vs.~$\varrho$ 
for different $(N-1) \rightarrow N$ transitions, with $\lambda=2$.
The dashed line represents the non-interacting orbital ($\lambda=0$). 
The ground states are $(M,S)=$ (0,1/2), (0,0), (1,1/2), (0,1),
(1,1/2), (0,0), for $N$ going from 1 to 6, respectively. 
$S_z=0$ ($S_z=1/2$) if $N$ is even (odd). The norm of the
$\lambda=2$ wavefunction is 1, 0.84, 0.84, 0.40, 0.37, 0.73, respectively.    
Insets: Total ground-state
charge densities $n(\varrho)$ (arb.~units)
for $N$ going from 1 (top left) to 6 (right bottom).
The length unit is $\ell_{\text{QD}}$.}
\label{figQP}
\end{figure}

\begin{figure}
\centerline{\epsfig{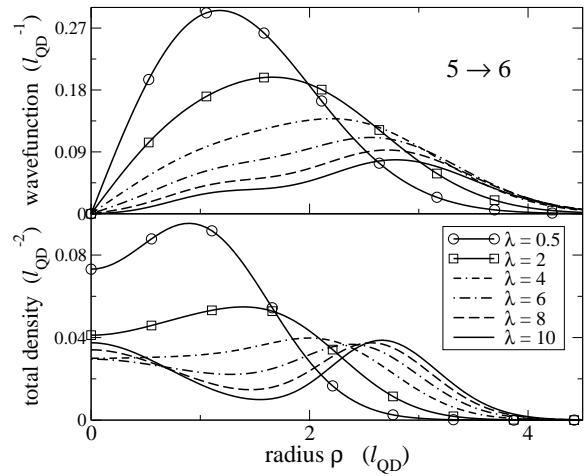}}
\caption{Top: Quasi-particle wavefunction vs.~$\varrho$ 
for different values of $\lambda$, as the 6-th electron tunnels into the QD. 
The wavefunction norm, for $\lambda$ going from 0.5 to 10, is
0.97, 0.73, 0.48, 0.32, 0.22, 0.15, respectively.
Bottom: Six-electron total density $n(\varrho)$
vs.~$\varrho$. The ground states for $N=5,6$ are $(M,S)=(1,1/2),(0,0)$,
respectively, for all $\lambda$.
}
\label{fig5-6}
\end{figure}

Figure \ref{figQP} shows $\varphi_{\text{QD}}$ vs.~$\varrho$, 
as up to six electrons are successively injected into a 
``liquid'' QD with a realistic 
density of $\lambda=2$ \cite{Seigo}. The QD filling sequence is well
known \cite{Ashoori,Seigo}, 
in analogy with the Aufbau principle of atomic physics:
in the independent-electron picture ($\lambda=0$, dashed lines),
$\varphi_{\text{QD}}$ is the highest-energy occupied orbital
which is filled by the electron added to the dot.
Howevever, Coulomb correlation significantly spreads the wavefunction
(solid lines) and moves the QP peak towards the QD edge.
The spreading is caused by the increase of weights $a_n$ of high-energy 
FD orbitals, as interaction is turned on; 
nevertheless, the behavior of $\varphi_{\text{QD}}$ around 
$\varrho\approx 0$ is dictated by its angular dependence,
$\varphi_{\text{QD}}(\varrho)\propto \varrho^{\left|m\right|}$,
while it decays like $\exp(-\varrho^2/2\ell^2_{\text{QD}})$
as $\varrho\rightarrow\infty$. 
The QP amplitude is strongly suppressed in the $(N-1)\rightarrow N$
tunneling processes involving the $N=4$
open-shell ground state, with respects
to other additions (Fig.~\ref{figQP}).
This is a spin-blockade effect,
since the total spin, $S$, is maximum at $N=4$
($S=1$ according to Hund's rule \cite{Seigo}), and we assume that its
$z$-component is zero, $S_z=0$. Besides, the general trend is that
the QP wavefunction norm, hence the integrated experimental signal, 
diminish as $N$ and $\lambda$ increase (see also Fig.~\ref{fig5-6}).

Note that the interpretation of tunneling spectroscopy in terms
of the total density, 
$n(\bm{\varrho})=\langle N | {\hat{\Psi}}^{\dagger}(\bm{\varrho})
\hat{\Psi}(\bm{\varrho})| N \rangle / N$, is inconsistent
with our point of view, as it is seen by comparing
QP wavefunctions of Fig.~\ref{figQP} with total densities for the
corresponding $N$-electron states (insets).
While total densities and QP probabilites  
resemble each other up to the addition of the second electron,
after the third electron tunnels into the dot they can be
clearly discriminated in the laboratory: 
QP probabilities have a strong angular dependence 
(hybridizing degenerate states with $\pm m$) and a node at
the QD center, while total densities are approximately circular
(exactly, for $N=4,6$) and filled.

As one reduces the density,
the appearance of QP wavefunctions dramatically changes.
In Fig.~\ref{fig5-6} we study the injection
of the 6th electron, as $\lambda$ goes from 0.5 up to 10.
Note that $\lambda=$ 0.5, 2, 4 (equivalent to GaAs lateral confinement 
energies $\hbar\omega_0=$ 47, 3.0, 0.74 meV, respectively),
typically correspond to different experimental QD
devices, such as self-assembled \cite{Patane,Fricke},
vertical mesa etched \cite{Seigo}, and 2DEG-depletion QDs
\cite{Ashoori}.
A six-electron Wigner molecule forms for $\lambda>4$, with one electron
localized at the QD center, and the remaining five 
arranged on an outer ring, at the vertices of a regular
pentagon \cite{Egger,Bolton,EPLus}. The crystallization
is clearly seen in the bottom panel of Fig.~\ref{fig5-6}, where,
as $\lambda$ increases, the total density develops one peak at $\varrho=0$,
for the central electron, and another one, close to
$\varrho=3\ell_{\text{QD}}$, for the outer ring.
Similarly, the five-electron molecule is a hollow 
pentagon \cite{Egger,Bolton}. In the top panel of Fig.~\ref{fig5-6}
we see that the QP wavefunction is strongly affected by
electron localization: while for $\lambda<4$ it somehow resembles
the non-interacting FD orbital $(n,m)=(0,-1)$, being spread
uniformly across the dot, for $\lambda>4$ it develops 
a well formed peak close to the outer-ring position.
The QP weight in the region inside the ring is strongly depleted,
eventually appearing as a shoulder of the main peak.
We conclude that, in the crystallized phase, the 6th electron can 
only enter the external ring, with negligible probability of
being located into the center. For smaller $N$, we find that electrons 
just enter the outer ring, since the pertinent geometrical
phases are hollow regular polygons \cite{complication}. 

We now come to the effect of a strong magnetic field
parallel to the tunneling direction $z$. 
As $B$ increases, the kinetic energy is quenched,
Landau bands of almost degenerate FD levels being formed.
$M$ increases due to Coulomb repulsion, since the higher $m$,
the outer the FD orbital \cite{Maksym}. In correspondence of ``magic''
values of $M$, the ground state turns out to be particularly
stable \cite{Maksym}: this family of incompressible \cite{Bob} states
has been varioulsy regarded as reminescent of fractional quantum Hall effect
(FQHE) states in two-dimensional electron layers \cite{Reimann},
or as a collection of Wigner molecules \cite{Maksymreview}.
  
In analogy with FQHE, it is convenient to introduce the filling factor 
$\nu$, defined as $\nu = N(N-1)/2M$, and to consider only FD levels in 
the lowest Landau band and full spin-polarization, 
which turns out to be a reasonable approximation 
at high $B$ \cite{Reimann}. In realistic situations, there are significant
$B$-ranges where $\nu$ is constant as $N$ is changed 
\cite{Reimann,Oosterkamp}. 
At $\nu = 1$, the interacting states are maximum density droplets
\cite{Reimann,MacDonald}, namely incompressible disks of almost uniform
density, \protect{$|N\rangle = \prod_{m=0}^{N-1}
{\hat{c}}^{\dagger}_{0m}|0\rangle$}, and $\varphi_{\text{QD}}$
is simply the highest occupied FD state, $\varphi_{0N-1}$, 
located at the edge of the dot, which is 
being filled by the tunneling electron, with $a_n=\delta_{n0}$:
\begin{equation}
\varphi_{\text{QD}}(\bm{\varrho})=\varphi_{n=0,m=N-1}(\bm{\varrho}).
\label{eq:MDD}
\end{equation}
Equation (\ref{eq:MDD}) is a remarkable result: while the total
electron density is a uniform disk, the measured squared modulus
of QP wavefunction will be an annulus of the same radius as the
charge distribution. 
If $\nu < 1$, the wavefunction will be still proportional
to the FD orbital $\varphi_{n=0,m}$, with $m = (N-1)/\nu$ 
and $a_0\neq 1$, namely
\protect{$\varphi_{\text{QD}}(\bm{\varrho})=a_0\,\varphi_{n=0,m}
(\bm{\varrho})$}.  The only effect of strong correlation in these
regimes is to modulate the amplitude of the non-interacting
wavefunction via the coefficient $a_0$. 
\begin{table}
\begin{ruledtabular}
\begin{tabular}{cccccc}
$(N-1)\rightarrow N$ 
& $\nu=1$ & $\nu=1/2$ & $\nu=1/3$ & $\nu=1/4$ & $\nu=1/5$ \\
\hline
$1\rightarrow 2$ & 1.00 &  1.00 & 0.500 & 0.707 & 0.250 \\
$2\rightarrow 3$ & 1.00 & 0.430 & 0.336 & 0.190 & 0.106 \\
$3\rightarrow 4$ & 1.00 & 0.520 & 0.270 & 0.201 & 0.0649 \\
$4\rightarrow 5$ & 1.00 & 0.158 & 0.239 & 0.0650 & 0.0507 \\
$5\rightarrow 6$ & 1.00 & 0.294 & 0.210 & 0.0541 & 0.0274 \\
\end{tabular}
\end{ruledtabular}
\caption{Absolute value of the modulation coefficient $\left|a_0\right|$ 
of the quasi-particle wavefunction 
\protect{$\varphi_{\text{QD}}(\bm{\varrho})=a_0\,\varphi_{0m}(\bm{\varrho})$},
where $m=(N-1)/\nu$, for different $(N-1)\rightarrow N$ tunneling
processes and filling factors $\nu$.
\label{tab1}}
\end{table}
Table \ref{tab1} shows the calculated values of
$a_0$ for various tunneling processes
at particularly stable filling factors. Except for some cases,
$\left|a_0\right|$ monotonously decreases as $\nu$ diminishes
or as $N$ increases. E.g., at $\nu=1/5$ $\left|a_0\right|$ is reduced by two
order of magnitudes with respect to $\nu=1$, when the 6th electron
enters the dot. Table \ref{tab1} shows that interaction enforces very 
effectively the orthogonality of incompressible states \cite{yannouleas}, 
and therefore we expect that, as a high field component is applied parallel 
to $z$, tunneling is strongly suppressed by the 
reduction of the matrix element $M_{\bm{k},N}$ [Eq.~(\ref{eq:M})]:
a purely many-body mechanism, the single-particle matrix element
$T_{\bm{k}}$ [Eq.~(\ref{eq:T})] being left unchanged by the field.

In conclusion, we have shown that quasi-particle wavefunctions of QDs
are extremely sensitive to electron-electron correlation, and may differ
from single-particle states in physically relevant cases. This result is 
of interest to predict the real- and reciprocal-space wavefunction images 
obtained by tunneling spectroscopies, as well as the intensities of 
addition spectra of QDs. Close comparison with experiment is not yet
possible in the case of Ref.~\cite{Wibbelhoff}, where many dots are
probed at once and the confinement is too strong. Promising samples
are also those of Refs.~\cite{Ashoori,Seigo}, allowing for access to a 
single dot and full control of $N$. We hope that our results will 
stimulate further experiments. We believe that our findings will be 
important also for other strongly confined systems, 
like e.g.~nanostructures at surfaces \cite{Leo}.

We thank O.~S.~Wibbelhoff and A.~Lorke for inspiring discussions
about their experiment. This paper
is supported by MIUR-FIRB RBAU01ZEML, MIUR-COFIN 2003020984,
I.T.~INFM Calc.~Par.~2004, MAE-DGPCC.

\clearpage
%
% Figures
%
%


\begin{thebibliography}{99}
\bibitem{orthogonality}
$\phi_{\bm{k}}$ and $\phi_{\alpha}$ are not mutually
orthogonal, since they are not eigenstates of the whole system Hamiltonian  
\cite{Bardeen}.
\bibitem{method}
For details on our CI approach see Ref.~\onlinecite{guy}. 
Here we implemented a parallel
version of our CI code, allowing for using a FD basis set at $B=0$
as large as 36 orbitals,
and for diagonalizing matrices of linear dimensions
up to $\approx 10^6$. As a convergence test, we could accurately 
reproduce Quantum Monte Carlo ground state energies up to $\lambda=8$ and 
$N=6$ \cite{Egger}. Details will be given in Ref.~\onlinecite{CI}.
\bibitem{Brueckner}
The dimensionless ratio $\lambda$ is the QD analog to the density parameter 
$r_s$ in extended systems.
\bibitem{complication}
There is a complication for the $3\rightarrow 4$ process,
since, for $\lambda$ between 4 and 6, 
the three-electron ground state becomes fully spin-polarized
\cite{Egger}, and the symmetry of the QP wavefunction changes
from $m=-1$ to $m=0$. Correspondingly, the $2\rightarrow 3$
channel is fully spin-blocked.
\end{thebibliography}
\end{document}